\documentclass[12pt]{article}

\usepackage{graphicx}
\begin{document}

\newcommand{\siml}{\stackrel{<}{\sim}}
\newcommand{\simg}{\stackrel{>}{\sim}}
\newcommand{\lleq}{\stackrel{<}{=}}



%
\begin{center}
{\large\bf
Information on mean firing rate, fluctuation and synchrony
conveyed by neuron ensembles with finite populations
\footnote{E-print: arXiv:0706.3489}
} 
\end{center}

\begin{center}
Hideo Hasegawa
\footnote{Electronic address: hideohasegawa@goo.jp}
\end{center}

\begin{center}
{\it Department of Physics, Tokyo Gakugei University  \\
Koganei, Tokyo 184-8501, Japan}
\end{center}
\begin{center}
({\today})
\end{center}
\thispagestyle{myheadings}

\begin{abstract}
A population of firing neurons is expected to carry 
not only mean firing rate but also its fluctuation 
and synchrony among neurons. In order to examine this possibility,
we have studied responses of neuronal ensembles
to three kinds of inputs: mean-, fluctuation- 
and synchrony-driven inputs.
The generalized rate-code model including additive and multiplicative 
noise (H. Hasegawa, Phys. Rev. E {\bf 75}, 051904 (2007))
has been studied by direct simulations (DSs)
and the augmented moment method (AMM) in which 
equations of motion for mean firing rate, fluctuation
and synchrony are derived.
Results calculated by the AMM are in good agreement
with those by DSs.
The independent component analysis 
(ICA) of our results has shown that mean firing rate, 
fluctuation (or variability) and synchrony may carry independent 
information in the population rate-code model.
The input-output relation of mean firing rates is shown 
to have higher sensitivity for larger multiplicative noise, 
as recently observed in prefrontal cortex.
A comparison is made between results obtained by the 
integrate-and-fire (IF) model and our rate-code model.
The relevance of our results to experimentally obtained
data is also discussed.

\end{abstract}

\noindent
\vspace{0.5cm}


\vspace{1cm}

\vspace{2cm}

\newpage

\section{Introduction}

One of the most important and difficult problems in neuroscience 
is to understand how neurons communicate in a brain.
There has been a long-standing controversy between
the temporal- and rate-code hypotheses in which information is
assumed to be encoded in firing timings and rates, respectively 
\cite{Rieke96}-\cite{deCharms00}.
A recent success in brain-machine interface (BMI)
\cite{Anderson04}\cite{Chapin99}, however, 
suggests that the population rate code is employed in
sensory and motor neurons, though it is still controversial
which of rate, temporal or
other codes is adopted in higher-level cortical neurons.

In recent years, much attention has been paid to 
a study on effects of mean firing rate, its fluctuation
and synchrony (or spatial correlation) of input signals
(for a review on rate and synchrony, see Ref. \cite{Salinas01}).
The precise role of synchrony in information transmission and
the relation among the firing rate, fluctuation and 
synchrony are not clear at the moment 
\cite{Salinas01}-\cite{Tiesinga04}.
The firing rate and synchrony are reported to be simultaneously  
modulated by different signals. For example, in motor tasks 
of monkey, firing rate and synchrony are considered to encode
behavioral events and cognitive events, respectively \cite{Riehle97}. 
During visual tasks, rate and synchrony are suggested to
encode task-related signals and expectation, respectively
\cite{Oliveira97}. A change in synchrony may amplify behaviorally 
relevant signals in V4 of monkey \cite{Fries01}.
An increase in synchrony of input signals is expected to yield 
an increase in output firing rate.
The synchrony of neurons in extrastriate visual cortex is, however,
reported to be modulated by selective attention even when there is
only small change in firing rate \cite{Tiesinga04}.
Rate-independent modulations in synchrony are linked to
expectation, attention and livalry \cite{Salinas01}.
Fluctuations of input signals have been reported to modify
the $f-I$ relation between an applied dc current $I$ and
autonomous firing frequency $f$ although its sensitivity to 
input fluctuation seems to depend on a kind of neurons 
\cite{Chance02}-\cite{Arsiero07}.
The $f-I$ curve of prefrontal cortex (PFC) retains the increased 
sensitivity to input fluctuations at large $I$, while that of 
somatosensory cortex (SSC) is insensitive to input fluctuation though
its linearity is increased at small $I$ \cite{Arsiero07}.

This kind of problems discussed above have been extensively studied 
by using spiking neuron models such as the Hodgkin-Huxley model 
\cite{Tiesinga00} and integrate-and-fire (IF) model with diffusion 
and mean-field approximations \cite{ReviewIF}-\cite{Heinzle07}.
The purpose of the present paper is to examine the same problem 
by using the rate-code model, which is an alternative theoretical 
model to the spiking model.
In a previous paper \cite{Hasegawa07b} (referred to as I hereafter), 
we proposed the generalized rate-code model for coupled
neuron ensembles with finite populations, which is subjected to 
additive and multiplicative noises.
It seems natural to include multiplicative noise beside additive 
noise in our rate-code model because the noise intensity is expected 
to generally depend on the state of neurons.
Actually effects of multiplicative noise in the spiking neuron model
are extensively examined by using conductance-based inputs which
yield multiplicative noise \cite{Rudolph03}-\cite{Richardson04}.
Our calculation in I has shown that the introduced multiplicative noise
leads to the non-Gaussian stationary distribution of firing rate, yielding
interspike-interval (ISI) distributions such as the gamma, 
inverse-Gaussian and log-normal distributions, which have been 
experimentally observed. We have discussed the dynamical properties 
of neuronal ensemble, by using the augmented moment method (AMM)
which was developed for a study of stochastic systems with finite 
populations \cite{Hasegawa03a}.
In the AMM, we pay our attention to global properties of neuronal 
ensembles, taking account of mean, and fluctuations of local and 
global variables.
The AMM has the same purpose to effectively study the properties 
of neuronal ensembles as approaches based on the population-code hypothesis
\cite{Knight00}-\cite{Rod96}. 
The AMM has been nicely applied to various subjects of neuronal ensembles 
\cite{AMM1}\cite{Hasegawa07c} and complex networks \cite{AMM2}. 

We have assumed in I that input signals are the same for all neurons
in the ensemble. In the present study, input signals are allowed 
to fluctuate and to be spatially correlated.
We will derive equations of motion 
for mean firing rate, its fluctuation and synchrony with the use of the AMM, 
in order to investigate the response to mean-, fluctuation-
and synchrony-driven inputs. This study clarifies, to some extent, 
their respective roles in information transmission.

The paper is organized as follows. In Sec. 2, we discuss an application 
of the AMM to the generalized rate model, studying the input-output 
relations of mean. fluctuation and synchrony. 
In Sec. 3, stationary and dynamical properties are discussed 
with numerical model calculations. By using the independent 
component analysis (ICA) \cite{Hyvarinen01}, we investigate 
a separation of signals when two or three kinds of inputs 
are simultaneously applied. Discussions are presented
in Sec. 4, where the results obtained by spiking IF model are
compared with those by our rate-code model.
The final Sec. 5 is devoted to conclusion. 

\section{Formulation}

\subsection{Adopted model}

For a study of the properties of a neuron ensemble containing 
finite $N$ neurons, we have adopted the generalized rate-code model 
\cite{Hasegawa07b,Hasegawa07c} in which a neuron is regarded
as a transducer from input to output signals, both of
which are expressed in terms of firing rates.
The dynamics of the firing 
rate $r_i(t)$ ($\geq 0$) of a neuron $i$ ($i=1$ to $N$) is given 
by 
\begin{eqnarray}
\frac{dr_{i}}{dt} &=& F(r_{i}) 
+H(u_{i})
+ G(r_{i}) \: \eta_{i}(t) + \xi_{i}(t), 
\label{eq:A1}
\end{eqnarray}
with
\begin{eqnarray}
u_{i}(t) &=& \left( \frac{w}{Z} \right) 
\sum_{j (\neq i)} \:r_{j}(t) + I_i(t), \label{eq:A2}\\
H(u) &=& \frac{u}{\sqrt{u^2+1}}\:\Theta(u). 
\label{eq:A3}
\end{eqnarray}
Here $F(r)$ and $G(r)$ are arbitrary functions of $r$, $Z$ $(=N-1)$ 
denotes the coordination number, $I_i(t)$ an input signal 
from external sources, $w$ the coupling strength, and $\Theta(u)$
the Heaviside function: $\Theta(u)=1$ for $u > 0$ and
$\Theta(u)=0$ otherwise.
Additive and multiplicative noises are included by $\xi_i(t)$ 
and $\eta_i(t)$, respectively, expressing zero-mean Gaussian white
noise with correlations given by
\begin{eqnarray}
\left< \eta_{i}(t)\:\eta_{j}(t') \right> 
&=& \alpha^2 \delta_{ij}\delta(t-t'), \label{eq:A4} \\
\left< \xi_{i}(t)\:\xi_{j}(t') \right> 
&=& \beta^2 \delta_{ij}\delta(t-t'), \label{eq:A5} \\
\left< \eta_{i}(t)\:\xi_{j}(t') \right> &=& 0,
\end{eqnarray}
where the bracket $\left< \cdot \right>$ denotes the average
over the distribution of $p(\{ r_i \},t)$ [Eq. (\ref{eq:X1})], and 
$\alpha$  and $\beta$ denote the magnitudes of multiplicative 
and additive noise, respectively.
The gain function $H(u)$ in Eq. (\ref{eq:A3}) expresses the response
of a rate output ($r$) to a rate input ($u$). 
It has been shown that, when spike inputs with mean firing rate $r_i$
are applied to a Hodgkin-Huxley (HH) neuron, mean firing rate $r_o$ of
output signals is $r_o \simeq r_i$ for $r_i \siml 60 $ [Hz], 
above which $r_o$ shows the saturation behavior \cite{Hasegawa00a,Wang06}.
The nonlinear, saturating behavior in $H(u)$ arises from the fact
that a neuron cannot fire with the rate of 
$r > 1/ \tau_r$ $(\equiv r_{max})$ where $\tau_r$ denotes the refractory 
period.
The function $H(u)$ has the rectifying property
because the firing rate is positive, which is expressed
by the Heaviside function in Eq. (\ref{eq:A3}).
Although our results to be present in the following are valid 
for any choice of $H(x)$, we have adopted, in this study, 
a simple analytic expression given by Eq. (\ref{eq:A3}), 
where the rate is normalized by $r_{max}$.

With the use of the diffusion-type approximation,
a spatially correlated input signal $I_i(t)$ 
in Eq. (\ref{eq:A2}) is assumed to be given by
\begin{eqnarray}
I_i(t) &=& \mu_I(t) + \delta I_i(t), 
\label{eq:A6}
\end{eqnarray}
with
\begin{eqnarray}
\langle \delta I_i(t) \rangle &=& 0, 
\label{eq:A7} \\
\langle \delta I_i(t) \delta I_j(t') \rangle 
&=&  [\delta_{ij} \gamma_I(t) + (1-\delta_{ij}) \zeta_I(t)]
\delta(t-t'),
\label{eq:A8}
\end{eqnarray}
where variance ($\gamma_I$) and covariance ($\zeta_I$) are given by
\begin{eqnarray}
\gamma_I(t) &=& \frac{1}{N} \sum_i \langle  \delta I_i(t)^2 \rangle, 
\label{eq:A9}\\
\zeta_I(t) &=& \frac{1}{NZ} \sum_i \sum_{j (\neq i)} 
\langle  \delta I_i(t) \delta I_j(t) \rangle.
\label{eq:A10}
\end{eqnarray}
We will discuss responses of the neuronal ensemble described 
by Eqs. (\ref{eq:A1})-(\ref{eq:A3}) to the spatially correlated input 
given by Eq. (\ref{eq:A6}), using both direct simulations (DSs) and
the AMM \cite{Hasegawa07b,Hasegawa03a}.

\subsection{Augmented moment method}

In the AMM \cite{Hasegawa03a}, we define the three quantities of 
$\mu$, $\gamma$ and $\rho$ given by
\begin{eqnarray}
\mu(t) &=& \langle R(t) \rangle 
= \frac{1}{N} \sum_i \langle r_i(t) \rangle, 
\label{eq:B1}\\
\gamma(t) &=& \frac{1}{N} \sum_i \langle [r_i(t)-\mu(t)]^2 \rangle, 
\label{eq:B2}\\
\rho(t) &=& \frac{1}{N^2} \sum_i \sum_j 
\langle [r_i(t)-\mu(t)][r_j(t)-\mu(t)] \rangle,
\label{eq:B3} 
\end{eqnarray}
where 
$R=(1/N) \sum_i r_i$,
$\mu(t)$ expresses the mean, and $\gamma(t)$ and $\rho(t)$
denote the averaged, auto and mutual correlations: 
related discussion will be given later [Eqs. (\ref{eq:E8})-(\ref{eq:E10})].

By using the Fokker-Planck equation (FPE), we get equations of motion for
$\langle r_i \rangle$ and $\langle r_i r_j \rangle$
(for details see appendix A). 
Expanding $r_i$ in Eqs. (\ref{eq:X2}) and (\ref{eq:X3}) around the average value 
of $\mu$ as
\begin{equation}
r_i=\mu+\delta r_i,
\label{eq:B4}
\end{equation}
and retaining up to the order of $ \langle \delta r_i\delta r_j \rangle$, 
we get equations of motion for $\mu$, $\gamma$ and $\rho$.
AMM equations in the Stratonovich representation are given by
(the argument $t$ being suppressed)
\begin{eqnarray}
\frac{d \mu}{dt}&=& f_{0}  + h_{0} + f_2 \gamma 
\nonumber \\
&+& \left( \frac{\alpha^2}{2}\right)
[g_{0}g_{1}+3(g_{1}g_{2}+g_{0}g_{3})\gamma], 
\label{eq:B6}\\
\frac{d \gamma}{dt} &=& 2f_{1} \gamma
+  \frac{2 h_{1} w}{Z}  \left(N \rho-\gamma  \right)
+ \gamma_I 
+ 2(g_{1}^2+2 g_{0}g_{2})\alpha^2\gamma
+ \alpha^2 g_{0}^2+\beta^2, 
\label{eq:B7}\\
\frac{d \rho}{dt} &=& 2 f_{1} \rho 
+ 2 h_1 w \rho
+ \frac{1}{N}(\gamma_I+ Z \zeta_I)
+ 2 (g_{1}^2+2 g_{0}g_{2}) \alpha^2 \rho
+\frac{1}{N}(\alpha^2 g_0^2+\beta^2), 
\label{eq:B8}
%
\end{eqnarray}
where $f_{\ell}=(1/\ell !)
(\partial^{\ell} F(\mu)/\partial x^{\ell})$,
$g_{\ell}=(1/\ell !)
(\partial^{\ell} G(\mu)/\partial x^{\ell})$, 
$h_{\ell}=(1/\ell !) 
(\partial^{\ell} H(u)/\partial u^{\ell})$ and
$u=w \mu + \mu_I$.
%
Original $N$-dimensional stochastic DEs given by 
Eqs. (\ref{eq:A1})-(\ref{eq:A3}) are
transformed to the three-dimensional deterministic DEs given 
by Eqs. (\ref{eq:B6})-(\ref{eq:B8}). 
For $\gamma_I=\zeta_I=0$, equations of motion 
given by Eqs. (\ref{eq:B6})-(\ref{eq:B8}) reduce to those obtained 
in our previous study \cite{Hasegawa07b}.
From $\mu$, $\gamma$ and $\rho$ obtained by Eqs.  (\ref{eq:B6})-(\ref{eq:B8}), 
we may calculate important quantities of synchrony and variability.
Then, they may be expressed in physically more transparent 
forms, as will be discussed in the following
[Eqs. (\ref{eq:D1})-(\ref{eq:D3})].

\subsection{Rate synchronization and variability}

\subsubsection{Synchronization ratio}

The synchronization is conventionally discussed for firing timings 
({\it temporal synchronization}) or phase ({\it phase synchronization}).   
We discuss, in this paper, the synchronization for firing rate
({\it rate synchronization}).
In order to quantitatively discuss the synchronization, we first 
consider the quantity $P(t)$ given by
\begin{equation}
P(t)=\frac{1}{N^2} \sum_{i j}<[r_{i}(t)-r_{j}(t)]^2>
=2 [\gamma(t)-\rho(t)].
\label{eq:C1}
\end{equation}
When all neurons are firing with the same rate
(the completely synchronous state),
we get $r_{i}(t)=R(t)$ for all $i$, and 
then $P(t)=0$ in Eq. (\ref{eq:C1}).
On the contrary, we get 
$P(t)=2(1-1/N)\gamma(t) \equiv P_{0}(t)$
in the asynchronous state where $\rho=\gamma/N$ 
\cite{Hasegawa07b,Hasegawa03a}.
We may define the normalized ratio for the synchrony of firing rates
given by \cite{Hasegawa03a}
\begin{equation}
S(t) \equiv 1-\frac{P(t)}{P_{0}(t)}
= \left( \frac{N \rho(t)/\gamma(t)-1}{N-1} \right)
= \frac{\zeta(t)}{\gamma(t)},
\label{eq:C2}
\end{equation}
where
\begin{eqnarray}
\zeta(t) 
&=& \frac{N}{Z}\left( \rho(t)-\frac{\gamma(t)}{N} \right), \\
&=& \frac{1}{N Z} \sum_i \sum_{j(\neq i)} 
\langle  \delta r_{i}(t) \delta r_{j}(t) \rangle.
\label{eq:C3}
\end{eqnarray}
$S(t)$ is 0 and 1 for completely asynchronous 
($P=P_0$) and synchronous states ($P=0$), respectively.

A similar analysis yields that the synchrony ratio of input signals 
may be expressed by
\begin{eqnarray}
S_I &=& \frac{(NZ)^{-1} 
\sum_i \sum_{j (\neq i)} \langle \delta I_i \delta I_j \rangle}
{N^{-1} \sum_i \langle (\delta I_i)^2 \rangle }
= \frac{\zeta_I}{\gamma_I},
\label{eq:C4}
\end{eqnarray}
where $\gamma_I$ and $\zeta_I$ are given 
by Eqs. (\ref{eq:A9}) and (\ref{eq:A10}),
respectively.

\subsubsection{Variability}

The variability in the ISI is usually defined by
by $C_V^T=\sqrt{\gamma^T}/\mu^T$ where $\mu^T$ and
$\gamma^T$ stand for mean and variance of ISI, respectively.
We here define the rate variability given by
\begin{eqnarray}
C_{V}(t) &=& \frac{\sqrt{\langle (\delta r_i)^2 \rangle}}{\mu}
= \frac{\sqrt{\gamma(t)}}{\mu(t)}.
\label{eq:C5}
\end{eqnarray}
Similarly, the variability in input rate signals is given by
\begin{eqnarray}
C_{VI}(t) &=& \frac{\sqrt{\langle (\delta I_i)^2 \rangle}}{\mu_I}
= \frac{\sqrt{\gamma_I(t)}}{\mu_I(t)}.
\label{eq:C6} 
\end{eqnarray}

Quantities calculated in the rate-code hypothesis 
are compared to those obtained in the temporal-code one
in the Appendix C.
We show in Eq. (\ref{eq:M17}) that if the temporal firing rate 
is given by $r_i(t)=1/T_i(t)$, 
we get $C_V(t) \simeq C_V^T(t)$, where $T_i(t)$ 
denotes the interspike interval (ISI) of firing times in a neuron $i$.

\subsection{AMM equations for $\mu$, $\gamma$ and $S$}

It is noted that the variability and synchrony are related to
the second-order statistics of local ($\gamma$) and global 
fluctuations ($\rho$), respectively, of firing rates.
Employing the relations given by Eqs. (\ref{eq:C2}) and (\ref{eq:C4}), 
we may transform equations of motion for $\mu$, $\gamma$ and $\rho$ given by 
Eqs. (\ref{eq:B6})-(\ref{eq:B8}) to those 
for $\mu$, $\gamma$ and $S$ given by
\begin{eqnarray}
\frac{d \mu}{dt}&=& f_{0} + h_{0} + f_2 \gamma 
+ \left( \frac{\alpha^2}{2}\right)
[g_{0}g_{1}+3(g_{1}g_{2}+g_{0}g_{3})\gamma], 
\label{eq:D1}\\
\frac{d \gamma}{dt} &=& 2f_{1} \gamma  
+ 2h_{1} w \gamma S
+ 2(g_{1}^2+2 g_{0}g_{2})\alpha^2\gamma
+ \gamma_I+ \alpha^2 g_{0}^2+ \beta^2, 
\label{eq:D2}\\
%
%
\frac{d S}{dt} 
&=&- \frac{1}{\gamma}(\zeta_I+ \alpha^2 g_0^2+\beta^2)S 
+ \left( \frac{\gamma_I S_I}{\gamma} \right)
+ \left( \frac{2 h_1 w}{Z} \right) (1+Z S)(1-S),
\label{eq:D3} 
\end{eqnarray}
where  
$f_{\ell}=(1/\ell !)
(\partial^{\ell} F(\mu)/\partial x^{\ell})$,
$g_{\ell}=(1/\ell !)
(\partial^{\ell} G(\mu)/\partial x^{\ell})$, 
$h_{\ell}=(1/\ell !) 
(\partial^{\ell} H(u)/\partial u^{\ell})$ 
with $u=w \mu + \mu_I$.
Equations (\ref{eq:D1})-(\ref{eq:D3}) express 
a response of $(\mu, \gamma, S)$ 
to a given input of $(\mu_I, \gamma_I, S_I)$.
Input signals in which information is encoded in
$\mu_{I}$, $\gamma_I$ and $S_{1}$ are hereafter referred to as
{\it mean-driven}, {\it fluctuation-driven} and {\it synchrony-driven} 
inputs, respectively.
Equations (\ref{eq:D1})-(\ref{eq:D3}) show that 
$\gamma_I$ plays a similar role
to independent noise while $\zeta_I$ ($=\gamma_I S_I$)
to correlated noise.

When we adopt $F(r)$ and $G(r)$ given by
\begin{eqnarray}
F(r)&=&-\lambda r^a, 
\label{eq:D4} \\
G(r)&=& r^b,
\label{eq:D5}
\hspace{2cm}\mbox{($a, b  \geq 0$)}
\end{eqnarray}
Eqs. (\ref{eq:D1})-(\ref{eq:D3}) become
\begin{eqnarray}
\frac{d \mu}{dt}&=&-\lambda \mu^a + h_0
-\left(\frac{\lambda}{2}\right) a(a-1)\mu^{a-2} \gamma 
\nonumber \\
&+& \left(\frac{\alpha^2}{2}\right)
[b \mu^{2b-1}+b(b-1)(2b-1)\mu^{2b-3} \gamma], 
\label{eq:D6}\\
\frac{d \gamma}{dt} &=& -2 \lambda a \mu^{a-1} \gamma 
+ 2 h_1 w \gamma S 
+ 2 b (2b-1) \alpha^2 \mu^{2b-2}\gamma 
+ \gamma_I + \alpha^2 \mu^{2b} + \beta^2, 
\label{eq:D7} \\
%
%
\frac{d S}{dt} &=& - \frac{1}{\gamma}
(2 c \gamma_I + \alpha^2 \mu^{2b}+\beta^2)S 
+\left( \frac{\gamma_I S_I}{\gamma} \right)
\nonumber \\
&+& \left( \frac{ 2 h_1 w}{Z} \right) (1+ZS)(1-S),
\label{eq:D8}
\end{eqnarray}
where $\lambda$ expresses the relaxation rate.
We note that for (i) $a=0$ or $1$ and (ii) $b=0$, $1/2$ or $1$,
a motion of $\mu$ is decoupled from the rest of variables because 
the $a(a-1)$ and $b(b-1)(2b-1)$ in the third and fourth terms of 
Eq. (\ref{eq:D6}) vanish. Equation (\ref{eq:D8}) shows that a motion of $S$ 
is ostensibly independent of the index $a$ of $F(r)$ although 
it depends on $a$ through $\gamma$.

\subsection{Relevance to experimental data}

Data obtained by neuronal experiments have been analyzed by using
various methods \cite{Salinas01}\cite{Brown04}-\cite{Averbeck06}.
It is worthwhile to discuss
the relevance of experimentally observed data to
$\gamma(t)$, $\rho(t)$, $\zeta(t)$ and $S(t)$ 
which are introduced in our study. 
Let $x_{i}^k(t)$ be a binned data defined by
\begin{eqnarray}
x_{i}^k(t) &=& 1,
\hspace{1cm}\mbox{if $t_{in}^k \in [t-\Delta t/2, t+\Delta t/2)$} 
\nonumber \\
&=& 0,
\label{eq:E1} \hspace{2cm}\mbox{otherwise}
\end{eqnarray}
where $t_{in}^k$ stands for the $n$th firing time 
of neuron $i$ of a given  $k$th trial
and $\Delta t$ the width of time window.
The firing rate of the neuron $i$ averaged 
over $M$ trials is given by
\begin{equation}
r_i(t) = \left( \frac{1}{M \Delta t} \right) 
\sum_{k=1}^{M} \:x_{i}^k(t).
\label{eq:E2}
\end{equation}
One conventionally calculates the correlation averaged over time
given by
\begin{eqnarray}
C_{ij}(\tau)&=&\int_{-\infty}^{\infty} \Gamma_{ij}(t,t+\tau)\:dt,
\label{eq:E3}
\end{eqnarray}
with
\begin{eqnarray}
\Gamma_{ij}(t,t')&=&
[r_{i}(t)-\mu(t)][r_{j}(t')-\mu(t')],
\label{eq:E4}
\end{eqnarray}
where 
$\mu(t)$ denotes the average of firing rates given by
\begin{equation}
\mu(t) = \frac{1}{N} \sum_{i=1}^{N} \:r_i(t).
\label{eq:E5}
\end{equation}
However, $C_{ij}(\tau)$ is not so useful for a study 
of the response to time-dependent inputs as in our case.
The equal-time, auto and mutual correlations
averaged over an ensemble are given by
\begin{eqnarray}
A(t) &=& \frac{1}{N} \sum_i \Gamma_{ii}(t,t), 
\label{eq:E6}\\
M(t) &=& \frac{1}{NZ} \sum_i \sum_{j(\neq i)} \Gamma_{ij}(t,t). 
\label{eq:E7}
\end{eqnarray}
Comparing Eqs. (\ref{eq:B2}) and (\ref{eq:C3}) with 
Eqs. (\ref{eq:E6}) and (\ref{eq:E7}), 
we get
\begin{eqnarray}
\gamma(t) &=& A(t), 
\label{eq:E8}\\
\zeta(t) &=& M(t),
\label{eq:E9}
\end{eqnarray}
which yields
\begin{eqnarray}
\rho(t) &=& \frac{1}{N} [A(t)+(N-1) M(t)], 
\label{eq:E10}\\
S(t) &=& \frac{M(t)}{A(t)}.
\label{eq:E11}
\end{eqnarray}
Equations (\ref{eq:E8})-(\ref{eq:E11}) show that 
$\gamma(t)$ and $\zeta(t)$ are equal-time,
auto and mutual correlations, respectively, and that
$S(t)$ is nothing but the 
mutual correlation normalized by the auto correlation.

\section{Model Calculations}

\subsection{Stationary properties}

In order to get an insight to the present method, we will show some 
model calculations. When we consider a special case of $a=b=1.0$ in
Eqs. (\ref{eq:D4}) and (\ref{eq:D5}):
\begin{eqnarray}
F(r)&=&-\lambda r, 
\label{eq:G1} \\
G(r)&=& r,
\label{eq:G2}
\end{eqnarray}
Eqs. (\ref{eq:D6})-(\ref{eq:D8}) are expressed  by
\begin{eqnarray}
\frac{d \mu}{dt}&=&-\lambda \mu + h_0
+ \frac{\alpha^2 \mu}{2}, 
\label{eq:G3}\\
\frac{d \gamma}{dt} &=& -2 \lambda \gamma 
+ 2 h_1 w \gamma S+ \gamma_I
+ 2 \alpha^2 \gamma  + \alpha^2 \mu^2 + \beta^2, 
\label{eq:G4}\\
%
%
%
\frac{d S}{dt} &=& - \frac{S}{\gamma} 
(2 c \gamma_I+\alpha^2 \mu^2+\beta^2)
+\frac{\gamma_I S_I}{\gamma}
+ \left( \frac{2 h_1 w}{Z} \right) (1+ZS)(1-S), 
\label{eq:G5}
\end{eqnarray}
The stationary solution of Eqs. (\ref{eq:G3})-(\ref{eq:G5}) is given by
\begin{eqnarray}
\mu &=& \frac{h_0}{(\lambda-\alpha^2/2)}, 
\label{eq:G6}\\
\gamma &=& \frac{(\gamma_I+\alpha^2\mu^2+\beta^2+2 h_1 w N\rho)}
{2(\lambda-\alpha^2+h_1 w/Z)}, 
\label{eq:G7}\\
%
%
S &=& \frac{Z(\lambda-\alpha^2)\gamma_I S_I
+h_1 w(\gamma_I+\alpha^2\mu^2+\beta^2)}
{(\gamma_I+\alpha^2\mu^2+\beta^2)
[Z(\lambda-\alpha^2)-h_1 w(Z-1)]+h_1 w \gamma_I S_I},
\label{eq:G8} \\
&=& \frac{\gamma_I S_I}{(\gamma_I+\alpha^2 \mu^2+\beta^2)},
\hspace{1cm}\mbox{for $w=0$}\\
&=& \frac{h_1 w}{[Z(\lambda-\alpha^2)-h_1 w (Z-1)]}.
\hspace{1cm}\mbox{for $S_I=0$}
\end{eqnarray}
We note in Eq. (\ref{eq:G6}) that $\mu$ is increased as $\mu_I$ is increased
with an enhancement factor of $1/(\lambda-\alpha^2/2)$,
but $\mu$ is independent of $\gamma_I$ and $S_I$.
Local fluctuation $\gamma$ is increased with increasing input fluctuation 
($\gamma_I$) and/or noise ($\alpha$, $\beta$) as Eq. (\ref{eq:G7}) shows.
Equation (\ref{eq:G8}) shows that $S$ is increased
with increasing $S_I$, as expected.

The stability condition around the stationary state may be examined
from eigenvalues of the Jacobian matrix of 
Eqs. (\ref{eq:G3})-(\ref{eq:G5}), which are 
given by (for details, see appendix B)
\begin{eqnarray}
\lambda_1 &=& -\lambda+\frac{\alpha^2}{2}+ h_1 w, \\
\lambda_2 &=& -2 \lambda+2 \alpha^2-\frac{2 h_1 w}{Z}, \\
\lambda_3 &=& -2 \lambda+2 \alpha^2+2 h_1 w. 
\end{eqnarray}
The first eigenvalue of $\lambda_1$ arises from an equation of 
motion for $\mu$, which is decoupled from the rest of variables. 
The stability condition for $\mu$ is given by
\begin{equation}
h_1 w < (\lambda-\alpha^2/2).
\label{eq:G11}
\end{equation}
In contrast, the stability condition for $\gamma$ and $\rho$ 
is given by
\begin{equation}
-Z (\lambda-\alpha^2) < h_1 w < (\lambda-\alpha^2).
\label{eq:G12}
\end{equation}
Then for $\lambda-\alpha^2 < h_1 w < \lambda -\alpha^2/2$,
$\gamma$ and $\rho$ are unstable but $\mu$ remains stable. 

The parameters in our model are $\alpha$, $\beta$, $w$ and $N$:
we hereafter set $\lambda=1.0$.
Input signals are characterized by $\mu_I$, $\gamma_I$ and $S_I$.
We will present some numerical calculations for
mean-, fluctuation- and synchrony-driven inputs applied to 
an ensemble with $\alpha=0.0$, $\beta=0.1$ and $N=100$ otherwise noticed. 

\vspace{0.5cm}
\noindent
{\bf A. Mean-driven inputs}

Figures 1 shows the $\mu_I$ dependences of $\mu$ and 
$S$ for $w=0.0$ (dashed curves) and $w=0.5$ (solid curves) 
with $\gamma_I=0.2$ and $S_I=0.2$. We note that for $w=0$, 
$\mu$ is increased with increasing $\mu_I$ after the gain function 
of $H(u)$, while $S$ is independent of $\mu_I$. 
In contrast, for finite $w$, $\mu$ is much increased than 
that for $w=0$. The chain line expressing 
$\mu=\mu_I$ crosses the $\mu$ curve at $\mu=0$ for $w=0.0$ and
at $\mu=0.735$ for $w=0.5$,

\vspace{0.5cm}
\noindent
{\bf B. Fluctuation-driven inputs}

Figure 2 shows $C_{V}$ and $S$ against
$C_{VI}$ ($=\sqrt{\gamma_I}/\mu_I$) for $w=0.0$ (dashed curves)
and $w=0.5$ (solid curves) with $\mu_I=0.2$ and $S_I=0.2$.
For $w=0.0$, an increase in $C_{VI}$ yields an increase in
$C_{V}$ and $S$, while $\mu$ is independent of $CV_I$, 
as Eqs. (\ref{eq:G6})-(\ref{eq:G8}) show. 
The chain curve shows $C_V=C_{VI}$:
the region where $C_V < C_{VI}$ is realized
for $C_{VI} > 0.51$ for $w=0.0$ and $C_{VI} > 0.22$ for $w=0.5$.

\vspace{0.5cm}
\noindent
{\bf C. Synchrony-driven inputs}

Figure 3 shows the $S_I$ dependences of $\mu$ and $S$ 
for $w=0.0$ (dashed curves) and $w=0.5$ (solid curves) 
with $\mu_I=0.2$ and $\gamma_I=0.2$.  
With increasing $S_I$, $S$ is increased as expected. 
For $w=0$, $\mu$ is independent of $S_I$,
as Eqs. (\ref{eq:G6}) shows. For $w=0.5$, $S$ is much increased 
compared to that for $w=0.0$.

It is necessary to point out that the $\mu_I$ dependence of $\mu$ is modified
by multiplicative noise ($\alpha$). 
An example of the $\mu_I$ dependence of $\mu$ is plotted in Fig. 4 
for various $\alpha$.
With increasing $\alpha$, $\mu$ shows a steeper increase for larger $\alpha$ 
because of the  $(\lambda -\alpha^2/2)$ factor in Eq. (\ref{eq:G6}).
This reminds us the recent experiment of prefrontal cortex (PFC)
showing that the $f-I$ curve has the increased sensitivity 
at large $I$ with increasing input fluctuation \cite{Arsiero07}.  
This is interpreted as due to a shorten, effective refractory
period by fluctuation in the calculation
using the IF model \cite{Arsiero07}.

The dependence of $S$ on $S_I$ is plotted in Fig. 5 for various values 
of $N$, ensemble size ($\alpha=0.5$, $\beta=0.2$, $w=0.5$).
It is shown that the synchrony $S$ is more increased in smaller system.
The result for $N=100$ is nearly the same as that for $N= \infty$.

\subsection{Dynamical properties}

\subsubsection{Pulse inputs}

In order to study the dynamical properties of the neuronal ensemble 
given by Eqs. (\ref{eq:A1})-(\ref{eq:A3}), we have performed direct simulations
(DSs) by using the Heun method \cite{Heun,Heun2} with a time step of 0.0001:
DS results are averages of 100 trials otherwise noticed.
AMM calculations have been performed 
for Eqs. (\ref{eq:G3})-(\ref{eq:G5}) by using the fourth-order Runge-Kutta method 
with a time step of 0.01. We consider, as a typical example, an ensemble
with $\lambda=1.0$, $\alpha=0.1$, $\beta=0.1$, $w=0.5$ and $N=100$.
Calculated responses to mean-, fluctuation- and
synchrony-driven pulse inputs are shown in Figs. 6(a), 6(b) and 6(c), where
solid and dashed curves show results of the AMM and DS, respectively.

\vspace{0.5cm}
\noindent
{\bf A. Mean-driven inputs}

First we apply a mean-driven pulse input given by
\begin{eqnarray}
\mu_I(t) &=& A \:\Theta(t-40) \Theta(60-t)+ A_b,
\label{eq:H1}
\end{eqnarray}
with $A=0.4$, $A_{b}=0.1$, 
$\gamma_I(t)=0.1$ and $S_I(t)=0.1$. 
Four panels in Fig. 6(a) show $\mu$, $\gamma$, $S$ and $C_{V}$:
chain curves express input signals
of $\mu_I(t)$, $\gamma_I(t)$ and $S_I(t)$.
An increase in an applied mean-driven input at $40 \leq t < 60$ induces 
an increase in $\mu(t)$ and decreases in $\gamma(t)$ and $S(t)$ 
which arise from the $h_1$ term 
in Eqs. (\ref{eq:G5}).
By an applied pulse input, $C_{V}(t)$ is decreased because of 
the increased $\mu$. The results of the AMM are in fairly good 
agreement with those of DS.

\vspace{0.5cm}
\noindent
{\bf B Fluctuation-driven inputs}

Next we apply a fluctuation-driven input:
\begin{equation}
\gamma_I(t)=  B \:\Theta(t-40) \Theta(60-t)+B_b,
\label{eq:H2}
\end{equation}
with $B=0.2$, $B_b=0.05$, 
$\mu_I(t)=0.1$ and $S_I(t)=0.1$, which are plotted by chain 
curves in Fig. 6(b). When the magnitude of $\gamma_I(t)$ is increased
at $40 \leq t < 60$, $\gamma(t)$ and $C_{V}(t)$ are much increased,
while there is no changes in $\mu(t)$.
$S(t)$ is modified only at $t \sim 40$ and $t \sim 60$, where the input 
pulse is on and off.

\vspace{0.5cm}
\noindent
{\bf c. Synchrony-driven inputs}

We apply a synchrony-driven input:
\begin{equation}
S_I(t)= C \:\Theta(t-40) \Theta(60-t)+C_b,
\label{eq:H3}
\end{equation}
with $C=0.4$, $C_b=0.1$,
$\mu_I(t)=0.1$ and $\gamma_I(t)=0.1$, which are plotted by 
chain curves in Fig. 6(c).
An increase in synchrony-driven input at $40 \leq t < 60$
induces increases in $S(t)$, $\gamma(t)$ and $C_{V}(t)$, 
but no changes in $\mu(t)$. 
This is because $\mu(t)$ is decoupled 
from the rest of variables in Eqs. (\ref{eq:G3})-(\ref{eq:G5}) 
for the case of $F(r)=-\lambda r$ and $G(r)=r$.

\subsubsection{Sinusoidal inputs}

We will apply mean-, fluctuation- and
synchrony-driven sinusoidal signals to a typical ensemble
with $\alpha=0.1$, $\beta=0.1$, $w=0.5$ and $N=100$.

\vspace{0.5cm}
\noindent
{\bf A. Mean-driven inputs}

First we apply a mean-driven sinusoidal input given by
\begin{eqnarray}
\mu_I(t)= 0.2 \: [1-\cos(2 \pi t/20)]+0.1,
\label{eq:H4}
\end{eqnarray}
with $\gamma_I(t)=0.1$ and $S_I(t)=0.1$.
Four panels in Fig. 7(a) show $\mu(t)$, $\gamma(t)$, $S(t)$ 
and $C_{V}(t)$ calculated by the AMM (solid curves) and DS 
(dashed curves): chain curves express input signals of 
$\mu_I(t)$, $\gamma_I(t)$ and $S_I(t)$.
An applied signal induces sinusoidal output in $\mu_I(t)$, and small 
changes in other quantities, as in the case of pulse inputs 
shown in Fig. 6(a).
  
\vspace{0.5cm}
\noindent
{\bf B. Fluctuation-driven inputs}

Next we apply fluctuation-driven sinusoidal inputs given by
\begin{eqnarray}
\gamma_I(t)= 0.05 \: [1-\cos(2 \pi t/20)],
\label{eq:H5}
\end{eqnarray}
with $\mu_I(t)=0.1$ and $S_I(t)=0.1$.
Calculated results are shown in Fig. 7(b), whose four panels show that 
an applied input induces sinusoidal changes in $\gamma(t)$ and $C_V(t)$,
but no changes in $\mu(t)$. $S(t)$ shows a peculiar time dependence
which may be compared to that in the case of pulse inputs shown
in the third frame of  Fig. 6(b). 

\vspace{0.5cm}
\noindent
{\bf B. Synchrony-driven inputs}

When we apply the synchrony-driven sinusoidal input given by
\begin{eqnarray}
S_I(t)= 0.25 \: [1-\cos(2 \pi t/20)],
\label{eq:H6}
\end{eqnarray}
with $\mu_I(t)=0.1$ and $\gamma_I(t)=0.1$. it leads to a significant 
change in $S(t)$ and also small changes in $\gamma(t)$ and $C_V(t)$, 
but no changes in $\mu(t)$, as in the case of pulse inputs
shown in Fig. 6(c).

\subsection{Independent component analysis}

It is interesting to estimate multivariate input signals
from multiple output signals.
Such a procedure has been provided in various methods
such as Bayesian estimation 
and independent component analysis (ICA) \cite{Hyvarinen01}.
Here we consider ICA, which was originally developed 
for a linear mixing system, and then it
has been extended to linear and nonlinear dynamical systems.
ICA has revealed many interesting applications in various fields 
such as biological signals and image processing.
A vector ${\sf x}$ of output signals is a real function ${\sf F}$
of a vector ${\sf s}$ of input sources:
\begin{equation}
{\sf x}={\sf F}({\sf s}).
\label{eq:J1}
\end{equation}
The dimension of ${\sf s}$ is assumed to be the same or
smaller than that of ${\sf x}$. If components of ${\sf s}$ are 
statistically independent and if only one of the source
signals is allowed to have a Gaussian distribution, ICA may extract 
a vector ${\sf y}$ with a function ${\sf G}$ given by
\begin{equation}
{\sf y}={\sf G}({\sf x}),
\label{eq:J2}
\end{equation}
from which we may estimate the original source as 
${\sf s} \simeq {\sf y}$ \cite{Hyvarinen01}.

\vspace{0.5cm}
\noindent
{\bf A. Coexistence of $\mu_I$ and $S_I$}

We will discuss the case when mean- and synchrony-driven inputs are
simultaneously applied to the neuronal model.
We consider the mean-driven sinusoidal input and synchrony-driven 
toothsaw input, given by
\begin{eqnarray}
\mu_I(t) &=& 0.1 \: [1-\cos(2 \pi t/20)]+ 0.1,
\label{eq:J3}\\
S_I(t) &=& 0.01 \:{\rm mod}(t,50),
\label{eq:J4}
\end{eqnarray}
with $\gamma_I(t)=0.1$ where mod($a,b$) denotes the mod function 
expressing the residue of $a$ divided by $b$.
Two panels in Fig. 8(a) show $\mu_I(t)$ and $S_I(t)$, and those 
in Fig. 8(b) show $\mu(t)$ and $S(t)$ calculated by the AMM.
We note a little distortion in $S(t)$ due to a cross talk from $\mu(t)$.
Assuming ${\sf s}=(\mu_I, S_I)^{\dagger}$ and ${\sf x}=(\mu, S)^{\dagger}$,
we have made an analysis of our result by using ICA.
Two panels in Fig. 8(c) show two components of ${\sf y}$ extracted 
from ${\sf x}=(\mu, S)^{\dagger}$ shown in Fig. 8(a) with the use of the 
fast ICA program \cite{ICA}. 
Although the program is designed for linear, mixing signals,
we have employed it for our qualitative discussion. 
We note that results in Fig. 8(c) fairly well reproduce 
the original, sinusoidal and toothsaw signals shown in Fig. 8(a).

\vspace{0.5cm}
\noindent
{\bf B. Coexistence of $\mu_I$, $\gamma_I$ and $S_I$}

Next we consider the case where three kinds of inputs are simultaneously 
applied. They are mean-driven sinusoidal signal,
fluctuation-driven toothsaw signal and synchrony-driven
square pulse signal, given by
\begin{eqnarray}
\mu_I(t) &=& 0.1 \: [1-\cos(2 \pi t/20)]+ 0.1,
\label{eq:J5}\\
\gamma_I(t) &=& 0.002 \:{\rm mod}(t,50), 
\label{eq:J6}\\
S_I(t) &=& 0.5 \:\Theta(-\cos(2 \pi t/120)).
\label{eq:J7}
\end{eqnarray}
Three panels in Fig. 9(a) show the input signals of
$\mu_I(t)$, $\gamma_I(t)$ and $S_I(t)$.
Output signals of $\mu(t)$, $\gamma(t)$ and $S(t)$ 
calculated in the AMM are shown in three panels of
Fig. 9(b). $\gamma(t)$ and $S(t)$ are a little distorted by a cross talk.
We have made an analysis of our result by using ICA, assuming
${\sf s}=(\mu_I, \gamma_I, S_I)^{\dagger}$ 
and ${\sf x}=(\mu, \gamma, S)^{\dagger}$.
Three panels of Fig. 9(c) show signals extracted by ICA. 
Extracted sinusoidal and square signals are similar to those
of input signals, though the fidelity of a toothsaw signal
is not satisfactory.
This is partly due to the fact that the fast ICA program adopted
in our analysis is developed for linear mixing models, but
not for dynamical nonlinear models \cite{ICA}.

These ICA analyses show that the mean rate, fluctuation and synchrony may
independently carry information in our population rate-code model.

\section{Discussion}
\subsection{A comparison with previous studies}

Various attempts have been proposed to obtain the firing-rate model, 
starting from spiking neuron models \cite{Amit91}-\cite{Aviel06}.
It is difficult to analytically calculate the firing rate
based on the firing model, except for the IF-type model
\cite{Brunel99}-\cite{Heinzle07}. 
In the coupled IF model, the dynamics of the membrane voltage 
$v_i(t)$ of the neuron $i$ ($=1-N$) is given by 
\begin{eqnarray}
\frac{dv_i}{dt} &=& -\frac{v_i}{\tau_m}  +I_i(t),
\label{eq:K1}
\end{eqnarray}
where $\tau_m$ denotes the relaxation time of the membrane and 
$I_i(t)$ stands for an input to neuron $i$.
When the mean-field and diffusion approximations are adopted, the 
input signal is given by \cite{Brunel99}
\begin{equation}
I_i(t)=J \mu_I(t)+ \beta(t) \xi_i(t),
\label{eq:K2}
\end{equation}
where $\beta(t)= \sqrt{J \gamma_I(t)}$,
$J$ is the all-to-all coupling, $\mu_I(t)$ and $\gamma_I(t)$ 
denote the mean and fluctuation, respectively, of input signals, and
$\xi_i(t)$ expresses zero-mean Gaussian white noise with correlations 
given by 
$\left< \xi_i(t)\:\xi_j(t') \right> = \delta_{ij}\:\delta(t-t')$. 
The firing of neuron is assumed to occur at $t=t_f$ when the 
voltage $v(t)$ crosses the threshold $\theta$ from below, 
and then the voltage is reset to the potential $v_r$:
$v(t_f)=\theta$, and $v(t_f+0)=v_r$.
The firing rate $r(t)$ is expressed by
\begin{equation}
r(t) = \frac{\beta(t)^2}{2}
\left[- \frac{\partial p(v,t)}{\partial v} \right]_{v=\theta},
\label{eq:K3}
\end{equation}
where the probability distribution $p(v,t)$ is calculated by the FPE 
with the boundary conditions at $v=\theta$ and $v=v_r$ and the 
normalization condition. For the stationary state, we get
\begin{eqnarray}
\frac{1}{r}
&=&\tau_r + \left( \frac{2}{\beta^2} \right)
\int_{v_r}^{\theta} \:du\: 
e^{\lambda_m (u-I/\lambda_m)^2/\beta^2}
\int_{-\infty}^{u}\:dv\:
e^{-\lambda_m (v -I/\lambda_m)^2/\beta^2},
\label{eq:K4}
\end{eqnarray}
where $I=J \mu_I$, $\lambda_m$=$1/\tau_m$ and $\tau_r$ stands 
for the refractory period.
In order to discuss the dynamics of firing rate, it is necessary 
to solve the FPE to get $p(v,t)$ by numerical methods, 
as was made in Ref. \cite{Lindner01}.
Equation (\ref{eq:K3}) shows that if information of input signal is encoded 
in fluctuation of $\gamma_I(t)$, its transmission is instantaneous, 
as experimentally observed \cite{Silberberg03,Lindner01}.

By adopting the IF model, Renart {\it et al.} \cite{Renart07}
have heuristically derived effective equations of motion for
the average firing rate $\mu_V$ and variance of the ISI $\sigma_V^2$ 
given by
\begin{eqnarray}
\frac{d \mu_V}{d t} &=& - \frac{1}{\tau_m}\mu_V + \mu_s, 
\label{eq:K5}\\
\frac{d \sigma_V^2}{d t} &=& - \frac{2}{\tau_m} \sigma_V^2
+ \sigma_s^2,
\label{eq:K6}
\end{eqnarray}
where $\mu_s$ and $\sigma_s$ are determined from the stationary
solution obtained from the FPE of the IF model [as given by Eq. (\ref{eq:K4})].
We note that Eqs. (\ref{eq:K5}) and (\ref{eq:K6}) are equivalent 
to Eqs. (\ref{eq:G3}) and (\ref{eq:G4}) 
for $\lambda = 1/\tau_m$ and $\alpha=0$.
Our AMM calculation provides us with not only equations of motion 
for mean and fluctuation but also that for synchrony.

Calculations with the use of the IF model have shown the followings: 

\noindent
(1) increased input firing rate deceases output variability 
\cite{Renart07},

\noindent
(2) increased input firing rate decreases synchrony \cite{Brunel00,Burkitt01}, 
\noindent

\noindent
(3) increased input fluctuation raises firing rate \cite{Arsiero07,Lindner03}, 

\noindent
(4) increased input synchrony increases firing rates
\cite{Salinas01,Tiesinga04,Shadlen98,Burkitt01,Moreno02}, and

\noindent
(5) increased input synchrony increases variability \cite{Salinas00}.

\noindent
The items (1), (2) and (5) are consistent with our result shown 
in Figs. 6(a) and 6(b). In contrast, items (3) and (4) seem to inconsistent 
with our result showing that $\mu(t)$ is independent of $\gamma_I(t)$ 
and $S_I(t)$, as given by Eqs. (\ref{eq:G3}).
It is noted that the model 
calculation given in the preceding section has been made for 
the case of $F(r)=-\lambda r$ and $G(r)=r$, in which a motion 
of $\mu(t)$ is decoupled from those of $\gamma(t)$ and $S(t)$.
This is not the case in general. 
We note in Eq. (\ref{eq:D6}) that $\mu(t)$ is not decoupled
from $\gamma(t)$ and $S(t)$ except for the case 
in which ($a=0$ or 1) and ($b=0$, $1/2$ or 1).  
For example, in the case of 
$F(r)=-\lambda r$ and $G(r)=r^2$, equations of motion given by
Eqs. (\ref{eq:D6})-(\ref{eq:D8}) become
\begin{eqnarray}
\frac{d \mu}{dt}&=&-\lambda \mu + h_0
+ \alpha^2 (\mu^{3}+ 3 \mu \gamma), 
\label{eq:L1} \\
\frac{d \gamma}{dt} &=& -2 \lambda \gamma 
+ 2 h_1 w \gamma S 
+ 12 \alpha^2 \mu^{2}\gamma 
+ \gamma_I + \alpha^2 \mu^{4} + \beta^2, 
\label{eq:L2} \\
\frac{d S}{dt} &=& - \frac{1}{\gamma}
(\gamma_I + \alpha^2 \mu^{4}+\beta^2)S 
+\frac{\gamma_I S_I}{\gamma}
\nonumber \\
&+& \left( \frac{2 h_1 w}{Z} \right) (1+ZS)(1-S),
\label{eq:L3}
\end{eqnarray}
Equations (\ref{eq:L1})-(\ref{eq:L3}) clearly show that $\mu(t)$ is coupled 
with $\gamma(t)$ and $S(t)$.
Figure 10(a), 10(b) and 10(c) show time courses of 
$\mu(t)$, $\gamma(t)$ and $S(t)$ given by 
Eqs. (\ref{eq:L1})-(\ref{eq:L3}) when mean-driven 
[Eq. (\ref{eq:H1}) with $A=0.4$, $A_b=0.1$], 
fluctuation-driven [Eq. (\ref{eq:H2}) with $B=0.2$, $B_b=0.05$] and
synchrony-driven pulse inputs [Eq. (\ref{eq:H3}) with $C=0.8$, $C_b=0.2$], 
respectively, 
are applied to a neuron ensemble with 
$\alpha=0.35$, $\beta=0.1$, $w=0.5$ and $N=100$.
We note in Fig. 10(b) that $\mu(t)$ is slightly 
increased when $\gamma_I(t)$ is increased
at $40 \leq t < 60$, while $\mu(t)$ is independent of $\gamma_I(t)$ 
for $F(r)=-\lambda r$ and $G(r)=r$ [Fig. 6(b)].
A similar increase in $\mu(t)$ is observed when synchrony-driven
input of $S_I(t)$ is applied as shown in Fig. 10(c),
while no changes in Fig. 6(c).
A peculiar time dependence is obtained in $\gamma(t)$  
for the mean-driven input in Fig. 10(a)  
compared to that in Fig. 6(a).
Except these points discussed above, responses in the case of $G(r)=r^2$ 
shown in Fig. 10 are similar to those in the case of $G(r)=r$ 
shown in Fig. 6. 

A number of neuronal experiments have not reported a systematic
changes in firing rates while the synchronization within 
an area is modulated \cite{Salinas01}. 
In particular, the synchrony is modified without a change 
in firing rate in some experiments
\cite{Riehle97,Fries01,Grammont03}.
It has been pointed out that such phenomenon may be
accounted for by a mechanism of a rapid activation
of a few selected interneurons \cite{Tiesinga04}.
Recently the absence of a change in firing rate is shown to be
explained if the ratio of excitatory to
inhibitory synaptic weights of long-range couplings is
kept constant in neuron ensembles described by the IF model
\cite{Heinzle07}.
It would be interesting to examine various cases
of $F(r)$ and $G(r)$, changing their
functional forms, which is left for our future study.

\section{Conclusion}

We have studied responses of neuronal ensembles to three kinds 
of inputs: mean-, fluctuation- and synchrony-driven 
inputs, applying the AMM and DSs to the generalized rate-code model 
\cite{Hasegawa07b}. The ICA analysis of our results
has suggested that mean rate, fluctuation (or variability) 
and synchrony may carry independent information. 
It would be interesting to examine this possibility 
by neuronal experiments using {\it in vivo} or {\it in vitro} 
neuron ensembles.

One of advantages of our rate-code model 
given by Eqs. (\ref{eq:A1})-(\ref{eq:A3}) is that we can easily
discuss various properties of neuronal ensembles, by changing
$F(r)$, $G(r)$ and $H(u)$, and model parameters.
We hope that our rate-code model shares advantages with 
phenomenological neuronal models such as the Hopfield \cite{Hopfield84} 
and Wilson-Cowan models \cite{Wilson72}. 
Although our rate-code model has no biological basis, 
it might be derived from spiking-neuron models with 
approaches previously adopted in Refs. \cite{Amit91}-\cite{Aviel06}. 
It is promising to apply the generalized rate-code model to
more realistic neuronal ensembles with excitatory and
inhibitory dynamical synapses.

\section*{Acknowledgments}
This work is partly supported by
a Grant-in-Aid for Scientific Research from the Japanese 
Ministry of Education, Culture, Sports, Science and Technology.  


\vspace{0.5cm}
\noindent
{\large\bf Appendix A: Derivation of AMM equations given by
Eqs. (\ref{eq:B6})-(\ref{eq:B8})}
\renewcommand{\theequation}{A\arabic{equation}}
\setcounter{equation}{0}

Using Eqs. (\ref{eq:A1}), (\ref{eq:A2}) and (\ref{eq:A6}), we get
\begin{eqnarray}
\frac{dr_{i}}{dt} &=& F(r_{i}) 
+H(u_{i}) + \delta I_i(t)
+ G(r_{i}) \: \eta_{i}(t) + \xi_{i}(t), 
\label{eq:X0}
\end{eqnarray}
with 
\begin{eqnarray}
u_{i}(t) &=& \left( \frac{w}{Z} \right) 
\sum_{j (\neq i)} \:r_{j}(t) + \mu_i(t). 
\end{eqnarray}
The Fokker-Planck equation for the distribution of
$p(\{r_{k} \},t)$ in the Stratonovich representation is given by 
\cite{Haken83,Ibanes99}
\begin{eqnarray}
\frac{\partial}{\partial t}\: p(\{r_{k} \},t)&=&
-\sum_{i} \frac{\partial}{\partial r_{i}}\{ [F(r_{i}) + H(u_{i})]
\:p(\{ r_{k} \},t)\}  
\nonumber \\
&+&\frac{1}{2}\sum_{i}\sum_{j}
\frac{\partial^2}{\partial r_{i} \partial r_{j}} 
\{ [ (\gamma_I+\beta^2)  \delta_{ij}+  \zeta_I(1-\delta_{ij}) ]
\:p(\{ r_{k} \},t) \},
\nonumber \\
&+&\frac{\alpha^2}{2}\sum_{i}
\frac{\partial}{\partial r_{i}} \{ G(r_i) 
\frac{\partial}{\partial r_i}
[ G(r_{i}) \:p(\{ r_{k} \},t) ] \}. 
\label{eq:X1}
\end{eqnarray}
Equations of motion for moments, $\langle r_{i}  \rangle$
and $\langle r_{i} \:r_{j} \rangle$, are derived with the use of FPE
\cite{Hasegawa07b}:
\begin{eqnarray}
\frac{d \langle r_i \rangle}{dt}
&=& \langle F(r_i) + H(u_i) \rangle
+\frac{\:\alpha^2}{2} \langle G'(r_i)G(r_i) \rangle,
\label{eq:X2}
\\
\frac{d \langle r_i \:r_j \rangle}{dt}
&=& \langle r_i\:[F(r_j)+H(u_j)] \rangle 
+ \langle r_j\: [F(r_i)+ H(u_i)] \rangle \nonumber \\
&+& \frac{\alpha^2}{2}
[\langle r_i G'(r_j) G(r_j) \rangle
+ \langle r_j G'(r_i) G(r_i)\rangle] \nonumber \\
&+& \delta_{ij} [\alpha^2\:\langle G(r_i)^2 \rangle +\gamma_I+\beta^2] 
+(1-\delta_{ij}) \zeta_I.
\label{eq:X3}
\end{eqnarray}

We may obtain Eqs. (\ref{eq:B6})-(\ref{eq:B8}), 
by using the expansion given by
\begin{eqnarray}
r_i &=& \mu + \delta r_i, 
\label{eq:X4}
\end{eqnarray}
and the relation given by
\begin{eqnarray}
\frac{d \mu}{dt} &=& \frac{1}{N} \sum_i 
\frac{d \langle r_i \rangle}{dt}, 
\label{eq:X6}
\\
\frac{d \gamma}{dt} &=& \frac{1}{N} \sum_i 
\frac{d \langle (\delta r_i)^2 \rangle}{dt},
\label{eq:X7} 
\\
\frac{d \rho}{dt} &=& \frac{1}{N^2} \sum_i \sum_j 
\frac{d \langle \delta r_i \delta r_j \rangle}{dt}.
\label{eq:X8}
\end{eqnarray}
For example, Eq. (\ref{eq:B6}) for $d \mu/dt$ is obtained as follows.
\begin{eqnarray}
\frac{1}{N} \sum_i \langle F(r_i) \rangle
&=& f_0+f_2 \gamma,
\label{eq:X9}\\
\frac{1}{N} \sum_i \langle H(u_i) \rangle
&=& h_0,
\label{eq:X10}
\\
\frac{1}{N} \sum_i \langle G'(r_i) G(r_i) \rangle
&=& g_0 g_1 + 3(g_0 g_3+g_1 g_2) \gamma.
\label{eq:X11}
\end{eqnarray}
Equations (\ref{eq:B7}) and (\ref{eq:B8}) are obtainable in a similar way.

\vspace{0.5cm}
\noindent
{\large\bf Appendix B: Jacobian matrix of Eqs. (\ref{eq:G3})-(\ref{eq:G5})}
\renewcommand{\theequation}{B\arabic{equation}}
\setcounter{equation}{0}

It is better to adopt the basis of $(\mu, \gamma, \rho)$ 
than to adopt that of $(\mu, \gamma, S)$, in making a linear stability 
analysis. In the former basis, equations of motion given
by Eqs. (\ref{eq:B6})-(\ref{eq:B8}) become 
for $F(r)=-\lambda r$ and $G(r)=r$:
\begin{eqnarray}
\frac{d \mu}{dt}&=&-\lambda \mu + h_0
+ \frac{\alpha^2 \mu}{2}, 
\label{eq:Y1}
\\
\frac{d \gamma}{dt} &=& -2 \lambda \gamma 
+ \frac{2 h_1 w N}{Z} \left( \rho-\frac{\gamma}{N} \right)
+ 2 \alpha^2 \gamma+ \gamma_I  + \alpha^2 \mu^2 + \beta^2, 
\label{eq:Y2}
\\
\frac{d \rho}{dt} &=& - 2 \lambda \rho 
+ 2 h_{1} w \rho + \frac{\gamma_I}{N}(1+ Z S_I) 
+ 2 \alpha^2 \:\rho +\frac{1}{N}(\alpha^2\mu^2+\beta^2),
\label{eq:Y3}
%
\end{eqnarray}

The stability of the stationary state may be examined by
the Jacobian matrix of Eqs. (B1)-(B3).
With the use of $c_{12}=c_{13}=c_{32}=0$ in the matrix, 
we get its eigenvalues given by
\begin{eqnarray}
\lambda_1 &=& c_{11}=-\lambda+\frac{\alpha^2}{2}+ h_1 w,
\label{eq:Y8} 
\\
\lambda_2 &=& c_{22}=-2 \lambda+2 \alpha^2-\frac{2 h_1 w}{Z}, 
\label{eq:Y9}
\\
\lambda_3 &=& c_{33}= -2 \lambda+2 \alpha^2+2 h_1 w,
\label{eq:Y10}
\end{eqnarray}
from which the stability condition given by
Eqs. (\ref{eq:G11}) and (\ref{eq:G12}) is obtained.

\vspace{0.5cm}
\noindent
{\large\bf Appendix C: Quantities in temporal and rate codes}
\renewcommand{\theequation}{C\arabic{equation}}
\setcounter{equation}{0}

We will discuss the relation between quantities calculated 
in the temporal code and those obtained in the rate code.
In the former, the interspike interval (ISI) defined by
\begin{eqnarray}
T_{i}^k(t) &=& t_{i n+1}^k-t_{i n}^k,
\hspace{1cm}\mbox{for $t_{in}^k \in [t-\Delta t/2, t+\Delta t/2)$},
\label{eq:M1}
\end{eqnarray}
plays an important role, where $t_{i n}^k$ denotes the $n$th 
firing time of neuron $i$ for a given $k$th trial.
When we assume that the firing rate is given by the inverse of ISI
averaged over $M$ trials:
\begin{equation}
\frac{1}{r_i(t)}= T_i(t) = \frac{1}{M} 
\sum_{k=1}^{M} T_i^k(t),
\label{eq:M2}
\end{equation}
the ISI histogram (ISIH) is given by
\begin{eqnarray}
P(T)=p(r) \mid \frac{dr}{dT} \mid
= \frac{1}{T^2}\: p\left(\frac{1}{T}\right),
\label{eq:M3}
\end{eqnarray}
where $p(r)$ denotes the distribution for firing rates.
For example, in the case of $F(r)=-\lambda r$,
$G(r)=r$ and $\beta=0$, we get stationary distributions
given by  \cite{Hasegawa07b}
\begin{eqnarray}
p(r) &\propto& r^{-(2 \lambda/\alpha^2+1)}
\exp\left[-\left(\frac{2H}{\alpha^2\:r}\right) \right], 
\label{eq:M4}\\
P(T) &\propto& T^{(2\lambda/\alpha^2-1)} 
\exp\left[-\left(\frac{2H}{\alpha^2} \right)T \right], 
\label{eq:M5}
\end{eqnarray}
where $P(T)$ is expressed by the gamma distribution.

The synchrony of ISI may be evaluated by its equal-time, 
auto and mutual correlations averaged over $N$ neurons and 
$M$ trials, as given by
\cite{Salinas01}\cite{Brown04}-\cite{Averbeck06}
\begin{eqnarray}
A^T(t)&=& \frac{1}{N} \sum_{i=1}^{N} 
\:[T_i(t)- \mu^T(t)]^2, 
\label{eq:M6}\\
M^T(t)&=& \frac{1}{N (N-1)} \sum_{i} \sum_{j (\neq i)}
[T_i(t)-\mu^T(t)][T_j(t)-\mu^T(t)],
\label{eq:M7}
\end{eqnarray}
with 
\begin{equation}
\mu^T(t)= \frac{1}{N} \sum_i \:T_i(t).
\label{eq:M8}
\end{equation}
A simple calculation using Eqs. (\ref{eq:M2}), (\ref{eq:M6}) 
and (\ref{eq:M7}) with the condition:
\begin{eqnarray}
&&\mid r_i - \mu(t) \mid \ll \mu(t), 
\label{eq:M9}\\
&&\mu(t) = \frac{1}{N} \sum_i \:r_i(t),
\label{eq:M10}
\end{eqnarray}
leads to
\begin{eqnarray}
A^T(t) & \simeq& \frac{A(t)}{\mu(t)^4}
=\frac{\gamma(t)}{\mu(t)^4}, 
\label{eq:M11}\\
M^T(t) & \simeq& \frac{M(t)}{\mu(t)^4}
= \frac{\zeta(t)}{\mu(t)^4},
\label{eq:M12}
\end{eqnarray}
where $A(t)$ and $M(t)$ denote auto and mutual correlations,
respectively, for firing rates given by Eqs. (\ref{eq:E6}) and (\ref{eq:E7}).
After Eq. (\ref{eq:E11}), we may define the normalized, temporal synchrony 
for ISI data, $S^T(t)$, given by
\begin{eqnarray}
S^T(t) &=& \frac{M^T(t)}{A^T(t)}.
\label{eq:M13}
\end{eqnarray}
Substituting Eqs. (\ref{eq:M11}) and (\ref{eq:M12}) to Eq. (\ref{eq:M13}), 
we get
\begin{equation}
S^T(t) \simeq S(t)=\frac{\zeta(t)}{\gamma(t)},
\label{eq:M14}
\end{equation}
where $S(t)$ stands for the normalized, rate synchronization
for firing rates given by Eq. (\ref{eq:C2}).

In order to assess the synchrony, one calculates the variance 
of ISI given by
\begin{eqnarray} 
C_V^T(t) &=& \frac{\sqrt{\gamma^T(t)}}{\mu^T(t)}, 
\label{eq:M15}\\
\gamma^T(t) &=& \frac{1}{N} \sum_i \:[T_i(t) -\mu^T(t)]^2.
\label{eq:M16}
\end{eqnarray}
From Eqs. (\ref{eq:M2}), (\ref{eq:M9}), (\ref{eq:M15}) and (\ref{eq:M16}), 
we get \cite{Hasegawa07b}
\begin{equation}
C^T_V(t) \simeq C_V(t)=\frac{\sqrt{\gamma(t)}}{\mu(t)},
\label{eq:M17}
\end{equation}
where $C_V(t)$ is the rate variance given by Eq. (\ref{eq:C5}). 
Stationary distributions given by Eqs. (\ref{eq:M4}) and (\ref{eq:M5}) yield
\begin{eqnarray}
C_V &=& \frac{\alpha}{\sqrt{2(\lambda-\alpha^2)}},
\label{eq:M18} \\
C_V^T &=& \frac{\alpha}{\sqrt{2 \lambda}},
\label{eq:M19}
\end{eqnarray}
which satisfy Eq. (\ref{eq:M17}) for $\alpha^2 \ll \lambda $.

According to neuronal experiments,
spike train variability is correlated with location in the 
processing hierarchy \cite{Harris05}. A large variability has 
been observed in hippocampus ($C_V^T \sim 3$) \cite{Fenton98}
and in visual V1 and MT of monkeys ($C_V^T = 0.5 \sim 1.0$) \cite{Softky92}
whereas a variation of motor neurons is very small 
($C_V^T = 0.05 \sim 0.1)$ \cite{Calvin68}.
For an explanation of the observed large $C_V^T$, several hypotheses
have been proposed: (1) a balance between excitatory and
inhibitory inputs \cite{Shadlen94}\cite{Troyer98}, 
(2) correlated fluctuations in recurrent networks \cite{Usher94},
(3) the active dendrite conductance \cite{Softky95},
and (4) a slowly decreasing tail of input ISI of
$T^{- d}$ ($d > 0$) at large $T$ \cite{Feng98}.
Equations (\ref{eq:M17})-(\ref{eq:M19}) show that the variance is increased 
with increasing multiplicative noise, which may be an alternative 
origin (or one of origins) of the observed large variability.

\newpage

\begin{figure}
\begin{center}
\end{center}
\caption{
(Color online)
Stationary values of $\mu$ $C_{V}$ and $S$
as a function of $\mu_I$ for $w=0.0$ (dashed curves)
and $w=0.5$ (solid curves) with
$\gamma_I=0.2$ and $S_I=0.1$
($\lambda=1.0$, $\alpha=0.0$, $\beta=0.1$, and $N=100$),
the chain curve expressing $\mu=\mu_I$.
}
\label{fig1}
\end{figure}

\begin{figure}
\begin{center}
\end{center}
\caption{
(Color online)
Stationary values of $C_{V}$ and $S$ 
as a function of $C_{VI}$ for $w=0.0$ (dashed curves)
and $w=0.5$ (solid curves) with 
$\mu_I=0.2$ and $S_I=0.2$
($\lambda=1.0$, $\alpha=0.0$, $\beta=0.1$, and $N=100$),
the chain curve expressing $C_{V}=C_{VI}$.
}
\label{fig2}
\end{figure}

\begin{figure}
\begin{center}
\end{center}
\caption{
(Color online)
Stationary values of $\mu$ and $S$
as a function of $S_I$ for $w=0.0$ (dashed curves)
and $w=0.5$ (solid curves) with 
$\mu_I=0.2$ and $\gamma_I=0.2$
($\lambda=1.0$, $\alpha=0.1$, $\beta=0.0$, and $N=100$),
the chain curve expressing $S=S_I$.
}
\label{fig3}
\end{figure}

\begin{figure}
\begin{center}
\end{center}
\caption{
$\mu$ as a function of $\mu_I$ for various
$\alpha$ values with $\lambda=1.0$ and $w=0.0$.
}
\label{fig4}
\end{figure}

\begin{figure}
\begin{center}
\end{center}
\caption{
$S$ as a function of $S_I$ for various
$N$ with  $\mu_I=0.1$, $\gamma_I=0.0$,
$\lambda=1.0$, $\alpha=0.5$, $\beta=0.2$ and $w=0.5$.
}
\label{fig5}
\end{figure}

\begin{figure}
\begin{center}
\end{center}
\caption{
(Color online)
(a) The time courses of $\mu(t)$, $\gamma(t)$,
$S(t)$ and $C_{V}(t)$ for mean-driven pulse input
given by Eq. (\ref{eq:H1}) with $A=0.4$, $A_{b}=0.1$, 
$\gamma_I=0.1$ and $S_I=0.1$:
(b) those for fluctuation-driven pulse input
given by Eq. (\ref{eq:H2}) with $B=0.2$, $B_b=0.05$,
$\mu_I=0.1$ and $S_I=0.1$:
(c) those for synchrony-driven pulse input
given by Eq. (\ref{eq:H3}) with $C=0.4$, $C_b=0.1$,
 $\mu_I=0.1$ and $\gamma_I=0.1$.
Solid and dotted curves express results of the AMM and DS,
respectively:
chain curves express inputs of
$\mu_I(t)$, $\gamma_I(t)$ and $S_I(t)$
($\lambda=1.0$, $\alpha=0.1$, $\beta=0.1$, $w=0.5$ and $N=100$).
}
\label{fig6}
\end{figure}

\begin{figure}
\begin{center}
\end{center}
\caption{
(Color online)
(a) The time courses of $\mu(t)$, $\gamma(t)$,
$S(t)$ and $C_{V}(t)$ for mean-driven sinusoidal input
given by Eq. (\ref{eq:H4}) with $\gamma_I=0.1$ and $S_I=0.1$:
(b) those for fluctuation-driven sinusoidal input
given by Eq. (\ref{eq:H5}) with $\mu_I=0.1$ and $S_I=0.0$:
(c) those for  synchrony-driven sinusoidal input
given by Eq. (\ref{eq:H6}) with $\mu_I=0.1$ and $\gamma_I=0.1$.
Solid and dashed curves express results of the AMM and DS,
respectively: chain curves express inputs of
$\mu_I(t)$, $\gamma_I(t)$ and $S_I(t)$
($\lambda=1.0$, $\alpha=0.1$, $\beta=0.1$, $w=0.5$ and $N=100$).
}
\label{fig7}
\end{figure}

\begin{figure}
\begin{center}
\end{center}
\caption{
ICA separation of the AMM result for 
mean- and synchrony-driven inputs:
(a) source input signals:
(b) output signals: 
(c) extracted signals by fast ICA \cite{ICA}
($\lambda=1.0$, $\alpha=0.1$, $\beta=0.1$, $w=0.5$ and $N=100$)(see text).
}
\label{fig8}
\end{figure}

\begin{figure}
\begin{center}
\end{center}
\caption{
ICA separation of the AMM result for 
mean-, fluctuation and synchrony-driven inputs:
(a) source input signals;
(b) output signals: 
(c) extracted signals by fast ICA \cite{ICA}
($\lambda=1.0$, $\alpha=0.1$, $\beta=0.1$, $w=0.5$ and $N=100$) (see text).
}
\label{fig9}
\end{figure}

\begin{figure}
\begin{center}
\end{center}
\caption{
(Color online)
(a) The time courses of $\mu(t)$, $\gamma(t)$,
$S(t)$ and $C_{V}(t)$ for mean-driven pulse input
given by Eq. (\ref{eq:H1}) with $A=0.4$, $A_b=0.1$,
$\gamma_I=0.2$ and $S_I=0.2$:
(b) those for fluctuation-driven pulse input
given by Eq. (\ref{eq:H2}) with $B=0.2$, $B_b=0.05$,
$\mu_I=0.2$ and $S_I=0.2$:
(c) those for synchrony-driven pulse input
given by Eq. (\ref{eq:H3}) with $C=0.8$, $C_b=0.2$,
 $\mu_I=0.3$ and $\gamma_I=0.2$.
Solid and chain curves express results of the AMM
and inputs signals, respectively, in the case
of $F(r)=-\lambda r$ and $G(r)=r^2$
($\lambda=1.0$, $\alpha=0.4$, $\beta=0.1$, $w=0.5$ and $N=100$)
(see text).
}
\label{fig10}
\end{figure}


\begin{thebibliography}{99}
\bibitem{Heun}
In the Heun method for the ordinary differential equation of 
$dx/dt=f(x)$,
a value of $x$ at $t+h$ is evaluated by
$x(t+h)=x(t)+(h/2)[f(x_0)+f(x_1)]$ with $x_0=x(t)$
and $x_1=x(t)+h f(x(t))$, while it is given by
$x(t+h)=x(t)+ h f(x(t))$ in the Euler method,
$h$ being the time step.
The Heum method for the stochastic ordinary differential equation
meets the Stratonovich calculus employed in the FPE 
\cite{Heun2}.
\end{thebibliography}
\end{document}